# Novel applications of Generative Adversarial Networks (GANs) in the analysis of ultrafast electron diffraction (UED) images

Hazem Daoud*,[1] Dhruv Sirohi*,[2] Endri Mjeku*,[3] John Feng,[1] Saeed Oghbaey,[1] and R. J. Dwayne Miller[4]
[1)] *Department of Physics, University of Toronto, Toronto, Ontario, M5S 1A7, Canada*[a)]
[2)] *Department of Engineering Science, University of Toronto, Toronto, Ontario, M5S 1A7, Canada*
[3)] *Department of Mathematics, University of Toronto, Toronto, Ontario, M5S 1A7, Canada*
[4)] *Departments of Physics and Chemistry, University of Toronto, Toronto, Ontario, M5S 3H6, Canada*

Inferring transient molecular structural dynamics from diffraction data is an ambiguous task that often requires different approximation methods. In this paper we present an attempt to tackle this problem using machine learning. While most recent applications of machine learning for the analysis of diffraction images apply only a single neural network to an experimental dataset and train it on the task of prediction, our approach utilizes an additional generator network trained on both synthetic data and experimental data. Our network converts experimental data into idealized diffraction patterns from which information is extracted via a convolutional neural network (CNN) trained on synthetic data only. We validate this approach on ultrafast electron diffraction (UED) data of bismuth samples undergoing thermalization upon excitation via 800 nm laser pulses. The network was able to predict transient temperatures with a deviation of less than 6% from analytically estimated values. Notably, this performance was achieved on a dataset of 408 images only. We believe employing this network in experimental settings where high volumes of visual data are collected, such as beam lines, could provide insights into the structural dynamics of different samples.

## I. INTRODUCTION

In recent years the study of neural networks (NNs)[1–3] has opened the way for an ever-growing list of applications,[4,5] owing to the exponential growth of computational power in modern processors compared to their predecessors. The digitization revolution has provided a constant stream of useful data to train NNs and enable advancements in algorithms that are more efficient computationally. These applications range from speech recognition[6] to autonomous driving[7] and medical diagnosis.[8] At the heart of deep learning lies the idea that any complex task that requires an input and an output can be modelled with a sufficiently complex function. The parameters of this function can be 'learned' using labelled data to produce the desired output. The architecture of the function and its training methods are the subject of ongoing research.[9] However, in all cases, the process of 'training' requires the minimization of some loss function. The loss function is designed to yield high loss if the predictions of the NN significantly deviate from the correct outputs of the labelled data. The loss is reduced as the NN predictions of the labelled data improve.

In chemistry and biology, scientists are often faced with molecules that have many degrees of freedom, yet there are only a few key modes that direct the chemical processes.[10] This reduction in dimensionality is governed by a complex matrix of forces between the individual atoms inside the molecule and there is no theoretical derivation for it so far.[11] This, in a way, resembles the minimization of a loss function in machine learning. Hence, one would naturally think that a machine learning algorithm that learns how this reduction in dimensionality takes place could provide significant insights into a theoretically intractable problem. One of the dream goals of time-resolved crystallography is to understand the dynamics along the key modes that govern chemical processes through time-resolved observation techniques.[12] Typically, short electron pulses[13], produced by DC and RF electric fields,[14–18], or x-rays[19,20] are used as probe pulses to study photochemical reaction pathways that are triggered through pump laser pulses. The obtained diffraction patterns are analyzed to extract information about the molecular dynamics.

However, one major obstacle is that it is impossible to directly invert reciprocal diffraction space into real space without more information, such as the signal field, due to the well-known phase problem.[21] Therefore, to create a molecular movie, a minimization function is usually used that combines the diffraction data with known information about the chemical structures involved in the phase transition.[22] This approach can be expanded upon to exploit deep learning algorithms to aid in a theoretically ambiguous task.

Previous studies demonstrated the usage of deep neural networks (DNNs), typically convolutional neural networks (CNNs), on x-ray diffraction images to classify space groups[23–27], extract features[28,29] and identify cer-

[a)]* These authors contributed equally to this work

tain materials.[30] However, one application that would be very useful in the field of ultrafast science is a full structural analysis of the molecular dynamics extracted from diffraction images in a time-resolved pump-probe experiment. In this paper, we take a step toward this goal. We present results of using DNNs, combining CNNs and generative adversarial networks (GANs), to analyze diffraction patterns obtained in an ultrafast electron diffraction (UED) experiment. Nanometer-thin layers of bismuth were deposited on 20 nm SiN windows to enable sufficient electron transmission to probe the structural transitions using 95 keV electron probes. The bismuth sample was excited with 800 nm laser pulses in the strongly driven limit for the solid-liquid phase transition. The effective time resolution for this instrument is on the 100 fs time scale.[31] Diffraction images obtained at various time delays between pump and probe and different laser fluences were used to train and test our NN. Our analysis method is generalizable and, thus, should serve as a step in a more general scheme to predict molecular behaviour at the ultrafast timescale through deep learning.

## II. STUDYING BISMUTH TRANSIENT TEMPERATURES USING NEURAL NETWORKS

Bismuth samples consisting of layers of various thicknesses on silicon wafers, processed to give an array of 20 nm thin SiN windows to act effectively as electron transparent windows, were prepared via sputtering. The samples were then subjected to 800 nm 100 fs laser pulses and probed with 95 keV electron pulses. The purpose of the study was to test the ultrafast dynamics of bismuth melting and compare it to previous studies.

The ultrathin bismuth metal sheets were excited sufficiently to exceed the solid-liquid phase transition. In comparison to the previous free standing bismuth studies[32] the effect of the SiN solid substrate support was to retard the onset of melting. A similar effect was observed using XFEL structural probes[33] if one compares fluences. The dynamics in this work were found to be much slower in the presence of a polymer substrate for the same fluence and required much higher fluence than the free standing Bi films to approach non-thermal melting dynamics. This point was not commented upon in this work as the focus was on developing a timing tool. The same effect of the substrate is clearly visible in the present data. In this case, the dynamics can be treated as a thermally driven process, which could be well modelled by the traditional Debye-Waller relations for lattice heating.

Hence, it is predicted that, at various time points in the initial picoseconds after laser excitation, the temperature of bismuth would vary by first rising and then stagnating at some point before cooling off.[32] The temperature can be theoretically estimated from the diffraction patterns directly via analytical calculations. However, it would be useful to build a NN that can predict the temperature from the diffraction patterns directly for several reasons:

1. This is a less complex problem that involves only temperature changes in one type of atom involved in electron scattering. Hence it can serve as a simpler test case as a step toward more complex problems.

2. Successful analysis of diffraction patterns via a NN architecture can be applied to other problems if trained with different data on the same or very similar NN architectures.

3. In diffraction experiments a large volume of data is collected, therefore, having a NN that can automatically sort it out and provide an initial assessment can save a lot of time and effort.

4. Generally, one of the bottlenecks of machine learning is providing enough labelled data for a NN to be trained to a satisfactory level. In the here presented approach, a key part of the network, namely a CNN that interprets images, is trained solely on labelled synthetic data. The process of creating, potentially unlimited, synthetic data for diffraction, if applied correctly, can be extended to many other problems.

## III. METHOD

To illustrate the difference between typical NNs containing only CNNs and our novel NN containing a CNN and a GAN and show the improvement brought about by GANs, we present both NNs and discuss their respective performances.

### A. Neural network with CNN

Synthetic diffraction patterns for bismuth at different temperatures were created using the programs CrystalMaker[34] and CrystalDiffract[34]. The initial conditions (electron energy = 100 keV, peak profile = Gaussian) for the simulations closely matched the experimental conditions. First, plots of radial powder diffraction patterns were created for different $u_{iso}$, where $u_{iso}$ is the root-mean-square (RMS) velocity of the atoms in the bismuth crystal. Then, using some simple code, which takes as input the radial intensity graph I = f(r) and simply plots this in polar coordinates (r, $\theta$) = f(r) to create an image of circular rings centred at the origin in the middle of the image, these plots were converted into ideal diffraction patterns, i.e., perfectly circular patterns without noise. Fig. 1 shows a graph of the plot and the images created for various values of $u_{iso}$. The images created were then modified to better resemble the experimental diffraction patterns obtained in the lab. Thus, a central diffraction spot, a synthetic beam stop modelled after the experimental one were added as well as a Gaussian background. Fig. 2 shows such images before and after modification. Overall, 2060 images were created in the $u_{iso}$ range of 0.0 - 1.02, with increments of 0.01, consisting of 20 images for each $u_{iso}$ value with varying amounts of artificial noise added. These images were split in a 90/10 ratio to create the training and validation set for the CNN, respectively.

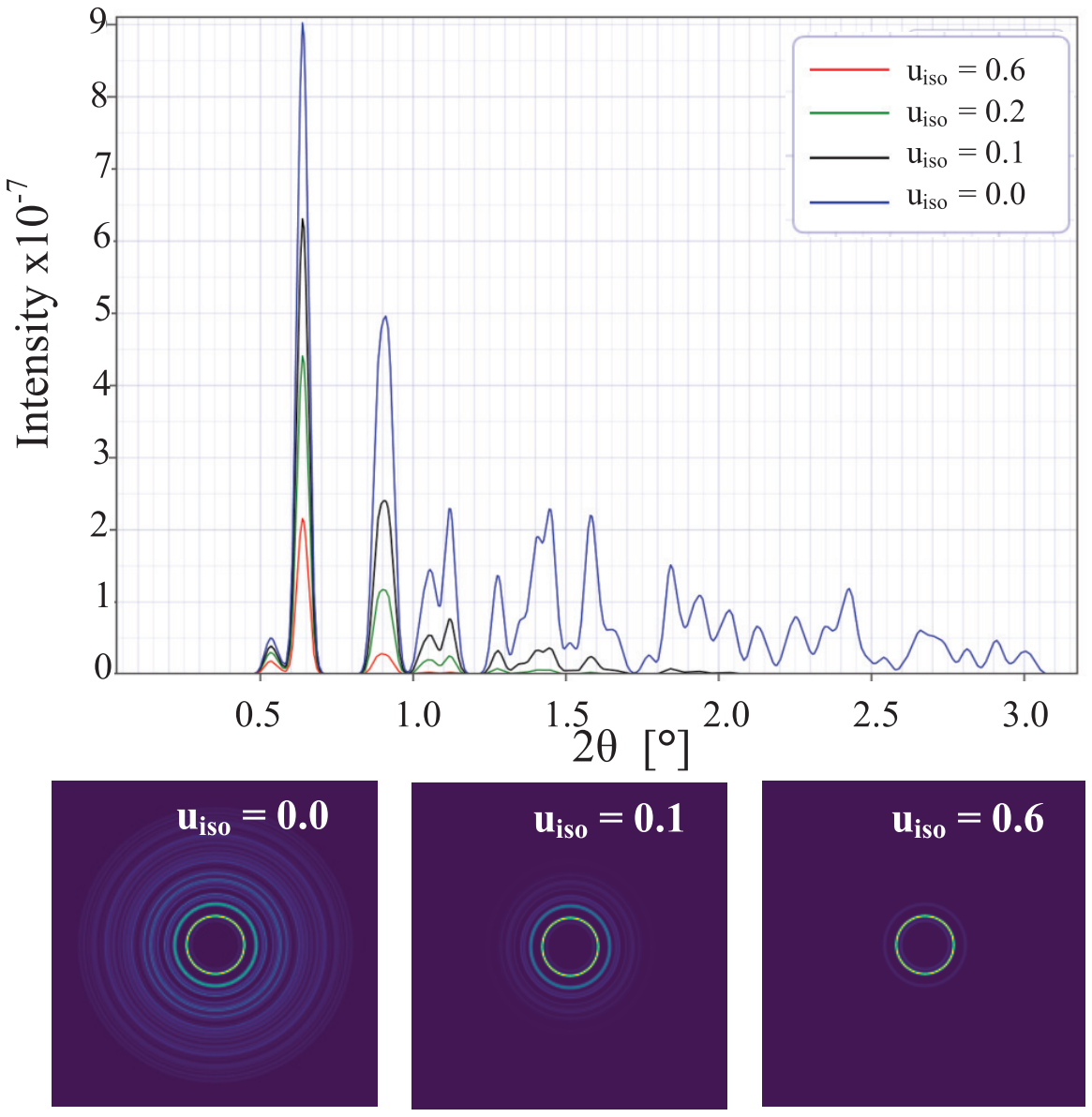

FIG. 1. Synthetically produced diffraction images. (Top) Radial intensity graphs generated via CrystalMaker for $u_{iso}$ values of 0.0, 0.1, 0.2, 0.6. (Bottom) The corresponding produced synthetic diffraction images.

### 1. Network architecture

We used a CNN consisting of three alternating convolutional and max-pooling layers followed by three dense layers and one dropout layer to reduce overfitting. Fig. 3 shows a schematic of the CNN. The images were reduced to 250 × 250 pixels and were fed into the network which contained the following layers: (a) Three convolutional layers that scan multiple filters through the image producing smaller images and highlighting certain features. (b) Three max-pooling layers that halve the dimension of the image by taking the brightest spot in 2 × 2 pixel grids across the image. (c) Three dense layers that flatten the image into a vector and apply an affine transformation to reduce each vector into a single number. (d) One dropout layer with a rate of 0.2 after the first dense layer, which drops out randomly 20% of the vectors produced by the previous layer on each training step. This reduces overfitting.

All convolutional and dense layers have rectified linear unit (ReLU) activation, which means that the outputs of these layers were passed through the function ReLu(x) = max(0, x), except for the last dense layer which had no activation function. The network was trained using a mean squared error loss, and the gradient descent algorithm used to minimize this loss was RMSprop. The learning rate of RMSprop was initialized to 0.001, but was set to exponentially decay throughout training with a decay rate of 0.96.

### 2. Main results and discussion

After 100 epochs of training, both the training and validation mean absolute errors reached below 0.03 as shown in Fig. 4.

Having trained the network on synthetically produced diffraction rings, we subsequently tested the accuracy of this network on real diffraction images obtained from the experiment. Electron diffraction images over a period of 20 ps and at various laser fluence levels were inputted into the trained NN. Fig. 5a shows a plot of the NN's predictions. As expected, the predicted temperatures before the laser pulse hit the sample (t = 0) are steady, but rise rapidly when the laser excites the sample. After a transient period the temperature reaches a steady state. In addition, we observe that, as the fluence level rises, the peak temperature also rises. This is expected as more energetic laser pulses cause more thermal motion in the bismuth atoms. Hence, the NN predicts the general trend successfully.

To quantitatively assess the accuracy of the network predictions, analytic methods to compute the true temperatures of the lab images were initially used, and then the NN $u_{iso}$ predictions were converted to temperature predictions, before comparing the two. Firstly, the temperature of the lab images was found using the equation[32]

$$B(T) = B(T_R) + \frac{2}{s^2} log(\frac{I_R}{I}) \quad (1)$$

to first find the Debye-Waller factor, which was then converted to temperature using the inverse of its fourth order polynomial approximation[35], computed via the Newton-Raphson method. I, $I_R$, T, $T_R$, and s denote the ring intensity, the ring intensity at room temperature $I_R$, the temperature, room temperature, and the scattering vector corresponding to the ring, respectively. Here, the intensity at room temperature was taken to be the intensity of the diffraction ring of the first image in each experiment, as each experiment started at room temperature, and the second diffraction ring was used, since the first was more prone to be obscured by noise.

To convert the NN's predicted $u_{iso}$ value to temperature, it was initially converted to the corresponding Debye-Waller factor B via the equation $B = 8\pi^2 u_{iso}$, before finally converting to temperature via the polynomial approximation mentioned previously.[35]

When the $u_{iso}$ levels are converted to temperatures (as seen in of Fig. 5b) a mismatch between experiment and predictions arises. From the analytic calculation using Equ. 1, temperatures below 1000 K are expected, but as can be seen from Fig. 5a, the NN predicts temperatures up to 2600 K. So, the network predicts higher temperatures systematically. Most likely, this is due to the experimental image being noisier than the ideal image. Additionally, the actual ratio of the intensity of some of the rings are different in the synthetic image compared to the lab images. Moreover, when training the CNN on a bigger dataset of 6080 synthetic images with temperatures of 293 K - 1540 K, which are much closer to the temperatures of the experimental images, the resultant predictions were also completely out of range, due the insufficient resemblance between the experimental and ideal images. The high accuracy of the NN on the synthetic images is expected to translate to high accuracy on the experimental images if the resemblance between the synthetic and experimental images was accurate enough.

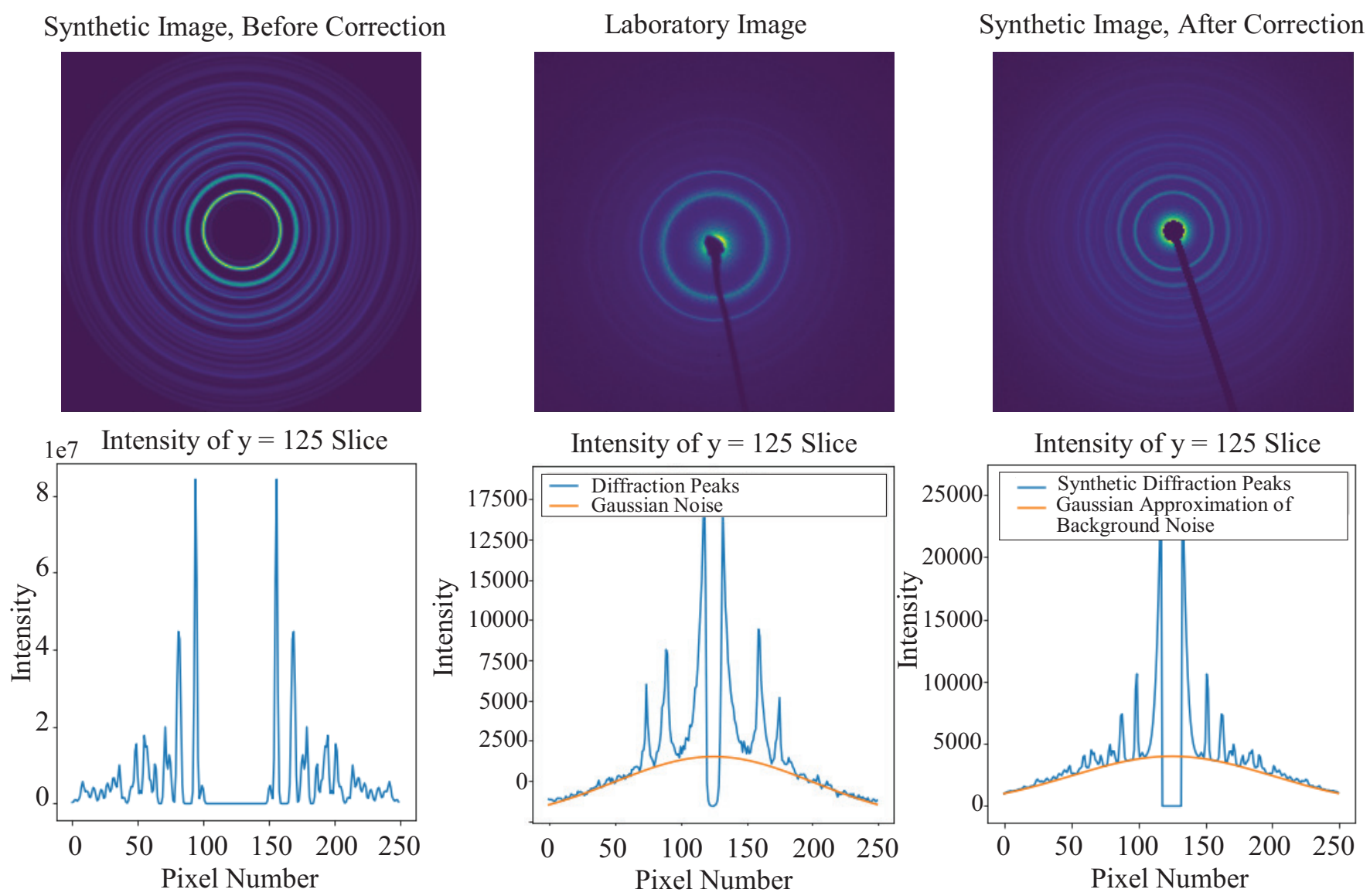

FIG. 2. Synthetic images after adjustment. (Top) Diffraction images. (Bottom) The corresponding horizontal intensity profile taken at the middle of the image.

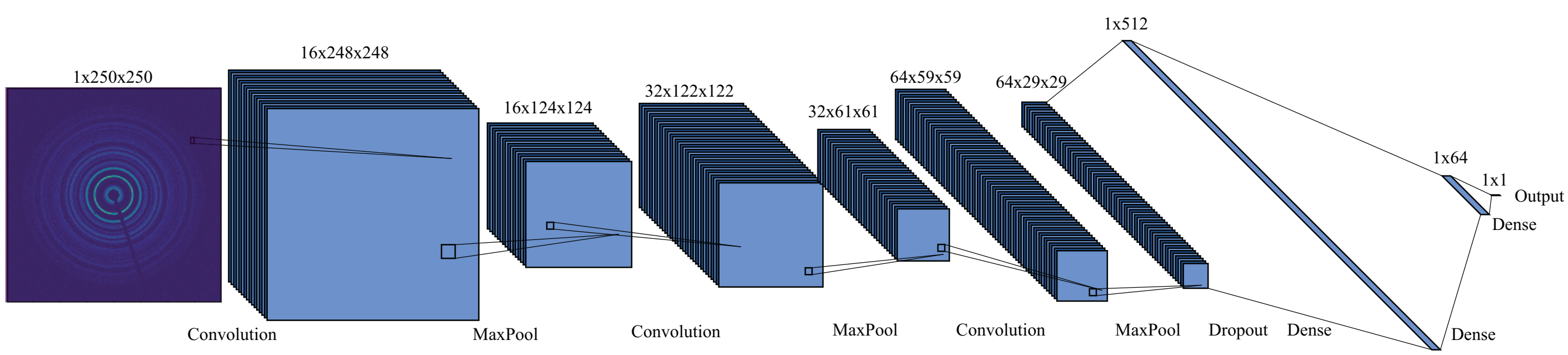

FIG. 3. A schematic of the CNN architecture.

To solve this issue, a different method of generating synthetic images was implemented. Since manually attempting to adjust the images did not yield quantitatively accurate results, a NN that outputs synthetic images that resemble the ideal diffraction patterns of experimental images was created. This would be beneficial as having labelled data to train NNs is often difficult to obtain. The ability to create labelled synthetic data would thus be very useful.

## B. GAN-CNN network

For the task of taking experimental diffraction images as input, converting them to ideal images similar to synthetic images created via CrystalDiffract and outputting a temperature value, we created a specific type of NN consisting of a GAN and a CNN.

### 1. GAN basic structure

A GAN consists of two NNs, a generator G and a discriminator D, training in tandem. The discriminator takes in a reference image and outputs the probability of

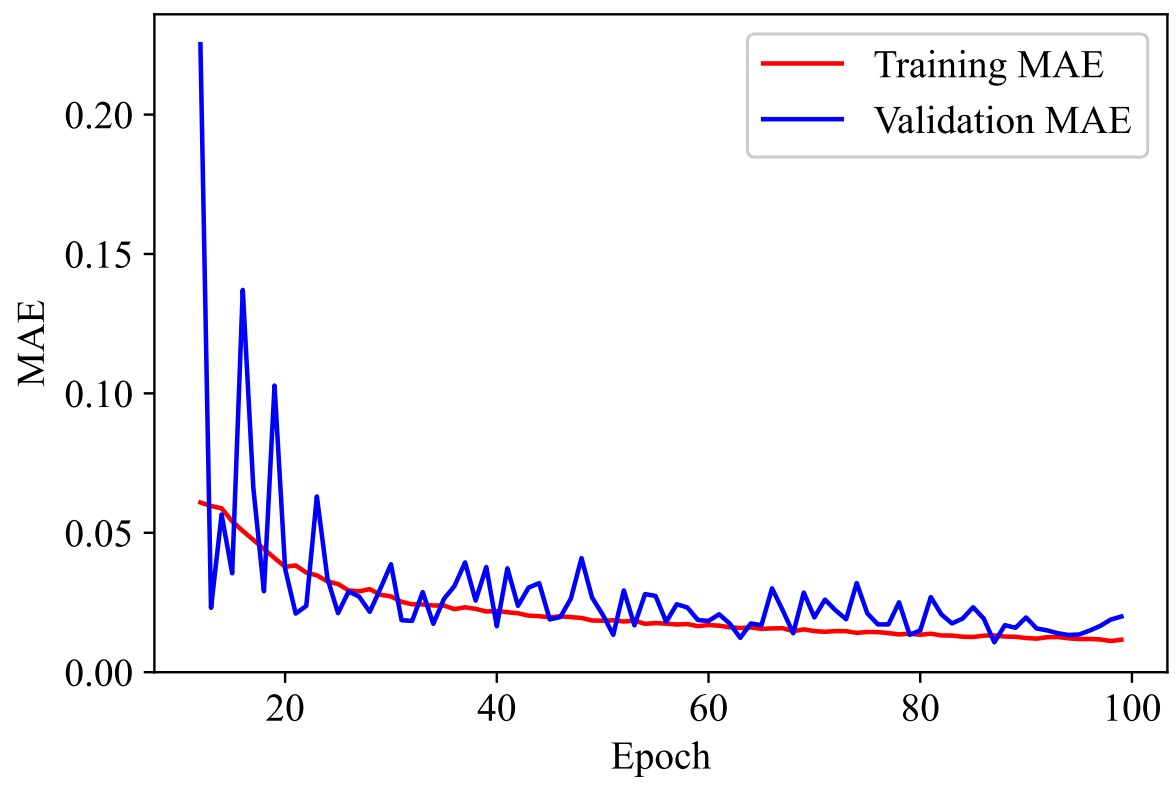

(a) Training and validation mean absolute error (MAE).

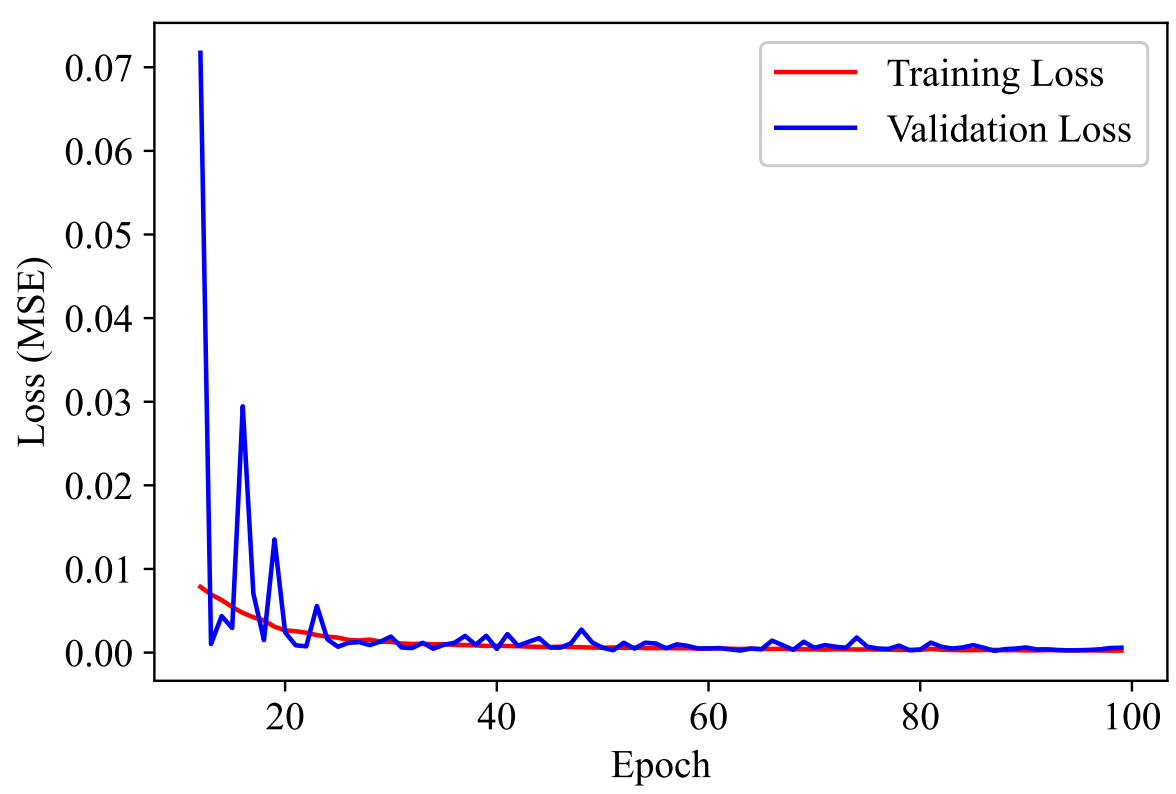

(b) Training and validation loss (mean squared error (MSE)).

FIG. 4. Training and validation accuracy and loss as a function of epochs.

it being a real image as opposed to a generated image. The generator takes in random noise and outputs generated images. The random noise input allows the generator to produce slightly different images for different noise vector inputs, hence introducing some variability in the output. The generator and discriminator compete against each other to raise each other's performance. As the discriminator gets better at recognizing 'fakes', the generator has to get better to continue 'fooling' it. The generator tries to create more accurate generated images that resemble the reference images to trick the discriminator, i.e., to maximize the probability of the discriminator incorrectly labelling generator images as real. The discriminator, on the other hand, tries to minimize the probability of it incorrectly detecting the generated images as reference images. Mathematically, the loss function of the GAN is

$$\mathcal{L}_{GAN}(G,D) = \mathbb{E}_{\mathbf{x}}[log(D(\mathbf{x})] + \mathbb{E}_{\mathbf{z}}[1 - log(D(G(\mathbf{z})))], \tag{2}$$

where $\mathbf{x}$ is the vector of reference data and $\mathbf{z}$ is the vector of random vectors. D($\mathbf{y}$) denotes the discriminator applied to the images $\mathbf{y}$, $\mathbb{E}_{\mathbf{y}}$ denotes the expected value over $\mathbf{y}$ and $G(\mathbf{z})$ denotes the generator applied to the random vector $\mathbf{z}$. During the training process the loss function is maximized for the discriminator and minimized for the generator:[36]

$$\min_{\mathrm{G}} \max_{\mathrm{D}} \mathcal{L}_{GAN}. \tag{3}$$

Maximizing this function corresponds to maximizing $D(\mathbf{x})$ and minimizing $D(G(\mathbf{z}))$ with respect to D, which corresponds to maximizing the probability of the discriminator identifying reference images as real and minimizing the probability of synthetic images being marked as real. This is how the discriminator improves its performance. On the opposite side, minimizing the loss function corresponds to maximizing $D(G(\mathbf{z}))$ with respect to G. This means improving the performance of the generator to produce images resembling the reference images enough to pass through the discriminator as real.

### 2. Network architecture

For our specific task we implemented a modified version of a GAN. While a basic GAN takes noise as input and outputs images, we implemented a GAN that takes experimental diffraction images and outputs ideal diffraction images similar to the ones synthetically created via CrystalDiffract. Experimental diffraction images typically include background noise due to scattering off the substrate holding the samples and dead or oversaturated pixels in the detector. This noise is typically removed with several analytical techniques, such as creating special masks to cover certain pixels or using specific equations to remove baseline noise,[37] and is not always accurate. Such a GAN then provides a mapping between experimental images and ideal synthetic images, of which an unlimited quantity can be created for training purposes.

Generating synthetic images given a lab image is an example of an image-to-image translation task, in which a mapping from an input image distribution to an output image distribution is to be learned. The data used for this task consists of image pairs. The model learns the mapping from the input image to the output image. If trained, the model can then be applied to new images to generate totally new outputs.

The dataset for image to image translation was created by pairing up lab images and synthetic images with the same temperature values. The model was trained to go from lab images to synthetic images. The synthetic images were cropped to 128 × 128 pixels and a Gaussian background was added in order to match the lab images as closely as possible, which led to better overall GAN performance. A circular mask was applied to the centre of both the lab and synthetic images. In total, 408 diffraction images were collected, of which 10% (41 images) were used as the validation set. Usually, image-to-image translation requires large datasets, often orders of magnitude larger than the size of the dataset available to us. Remarkably, despite the important differences between the images being minute our GAN is still able to recognize them. Fig. 6 shows plots of an experimental image, the corresponding generated image and the corresponding synthetic image created via CrystalDiffract.

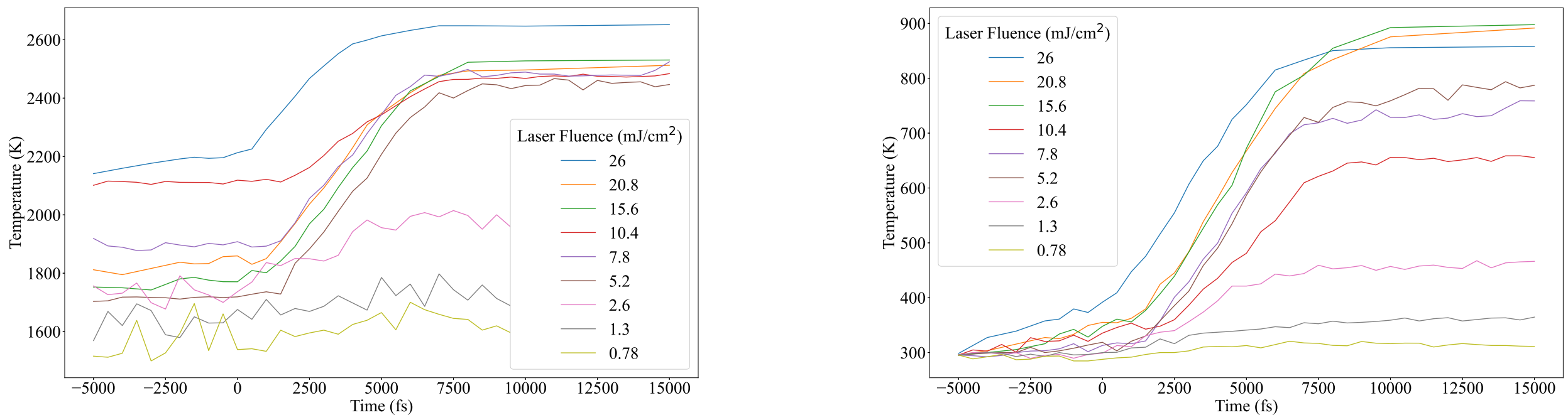

(a) CNN predicted temperatures.

(b) Analytically calculated temperatures.

FIG. 5. The neural network's prediction of the $u_{iso}$ values and temperatures of the experimental bismuth diffraction images. The laser pulse excites the sample at t = 0 fs, where a rapid rise in temperature is observed. The temperatures predicted by the CNN are inaccurate albeit they follow the general trend of rising after excitation.

For this task, we used a modified version of the Pix2Pix Generative Adversarial Network architecture.[38] Pix2Pix is built on the U-Net architecture, which has 3 components: a decoder, an encoder, and skip-connections. The decoder greatly down-samples the model's input until it reaches a bottleneck layer. The bottleneck layer is the layer with the fewest nodes in the network; placing it between the decoder and encoder forces the decoder to only retain essential information. The encoder upsamples this to get the model's output. However, there is often some low-level information that may be lost in the down-sampling process. To reliably translate the input image to the output, these features may be needed. The U-Net does this by incorporating skip-connections that connect the $i^{th}$ decoding layer from the input to the $i^{th}$ encoding layer from the output. For example, there would be a skip connection from the first decoder layer to the last encoder layer. Information can bypass the bottleneck layer by passing through these skip connections. The Pix2Pix loss function we used is similar to the GAN loss function described previously, with the addition of a loss term that measures the squared Euclidean distance between the pixel values of the generated images and the pixel values of the corresponding synthetic images.

In order to further increase performance on this task, further modifications were made to this loss function. Peak-signal-to-noise ratio (PSNR) and structural similarity index measure (SSIM) values between the generated images and synthetic images were also calculated.[39] These terms were included in the loss function to penalize generated images that are dissimilar to the corresponding synthetic images. These three terms, along with the GAN loss, are primarily effective at the start of training, when the GAN learns the general structure of the output images. On their own, however, they are not sufficient for the task: the network fails to consistently learn some small details of the synthetic images.

A new loss term is introduced to address this. This loss term is calculated by using a new CNN identical to the network in section III A. This Conversion CNN (CCNN) is trained to predict the $u_{iso}$ of synthetic images. It takes a generated image as input and outputs a corresponding $u_{iso}$ value. The difference between this predicted $u_{iso}$ and the analytically calculated $u_{iso}$ of the lab image is added as the new loss described in Equ. 4.

$$\mathcal{L}_{ImageTranslation}(\mathbf{x},\mathbf{y},\mathbf{u_{iso}}) = \alpha_1\mathcal{L}_{GAN}(G,D) + \beta_1 PSNR(G(\mathbf{x}),\mathbf{y}) + \beta_2 SSIM(G(\mathbf{x}),\mathbf{y}) + \gamma_1|\mathbf{y}-G(\mathbf{x})|_2^2 + \gamma_2|\mathbf{u_{iso}} - CCNN(G(\mathbf{x}))|. \quad (4)$$

Here, $\alpha_1$, $\beta_1$, $\beta_2$, $\gamma_1$ and $\gamma_2$ are manually-tuned hyperparameters, $\mathbf{x}$ is the vector of lab images, $\mathbf{y}$ is the vector of corresponding synthetic images created via CrystalDiffract, and $\mathbf{u_{iso}}$ is the vector of corresponding $u_{iso}$ values.

The GAN outputs a generated image that emulates an ideal diffraction pattern. This generated image is used as input for the CCNN, which outputs the corresponding $u_{iso}$ value. This predicted $u_{iso}$ value is the final output of the entire NN. It is compared to the analytically calculated $u_{iso}$ value, for images in the validation set, to quantify the percentage error of the NN. Fig. 7 shows a schematic of how the network operates to predict temperatures.

Fig. 8 shows a schematic of the training process of the NN. The generator takes a lab image as input and generates an image emulating an ideal diffraction pattern. The discriminator takes a generated image and a synthetic image containing an ideal diffraction pattern, with minimal processing, created via CrystalDiffract and outputs a value between 0 and 1 from which the discrimi-nator loss is calculated. The total loss function of the NN is the sum of the squared Euclidean distance between the generated image and the synthetic image, the difference between the analytical value of $u_{iso}$ and the value of $u_{iso}$ predicted by the CCNN, SSIM and PSNR of the generated image and the conventional generator loss which includes the discriminator predictions.

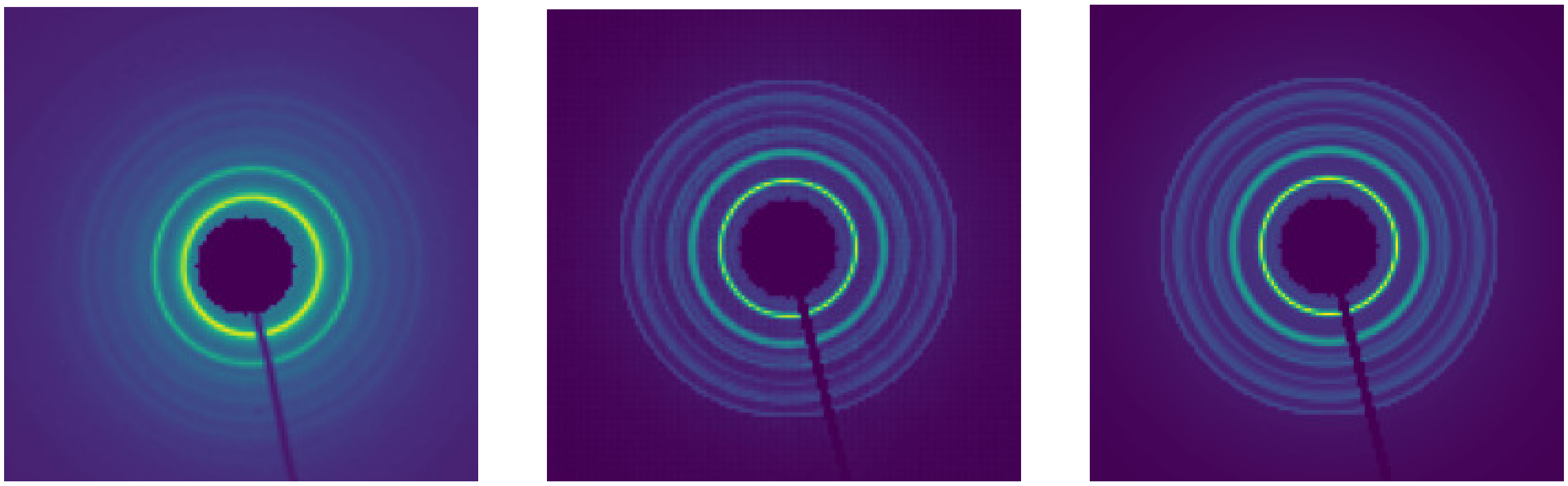

(a) Lab Image. (b) Generated Image. (c) Synthetic Image.

FIG. 6. Diffraction images: lab image, generated image, synthetic image. The generator takes a lab image as input and is trained to output the corresponding ideal diffraction pattern emulating the synthetic image.

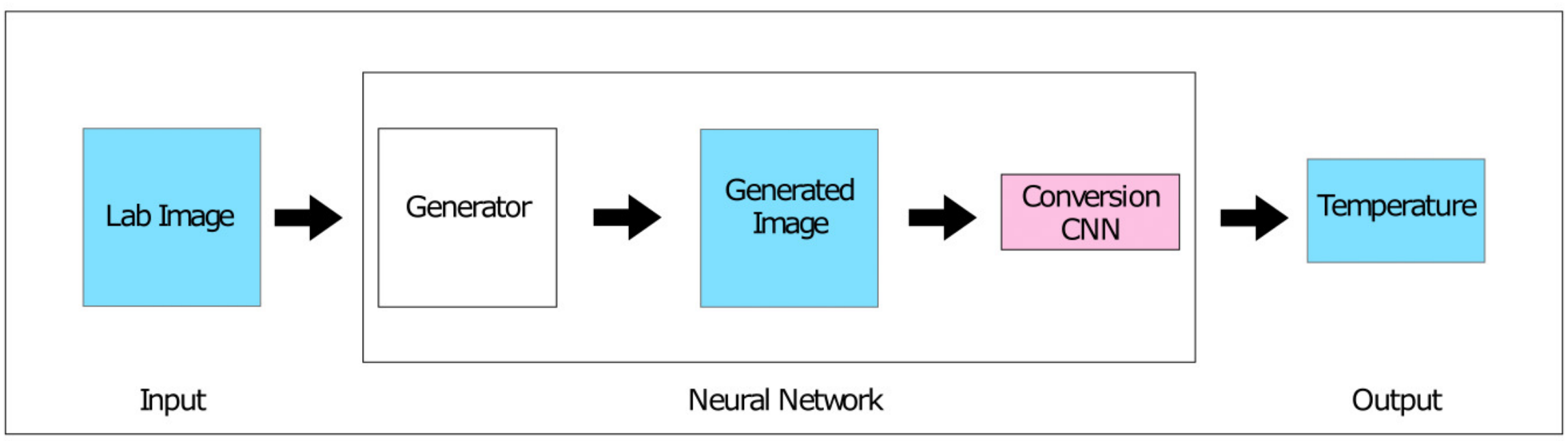

FIG. 7. GAN-CNN network operation schematic. An image containing a diffraction pattern is used as input to the generator, which generates an ideal diffraction pattern that is then converted via the Conversion CNN to a $u_{iso}$ value corresponding to a temperature.

### 3. Main results and discussion

After 18 epochs of training, a prediction error of 4.3% was reached on the training set and 5.72% on the validation set. Fig. 9 shows the training and validation errors. These results vary by a few percentage points for different training runs. This is due to some randomness in the initialization of the generator and discriminator networks at the start of the training process. A k-fold cross validation, with k = 5, yielded a prediction error of 9.23%.

This is a significant improvement over the performance in section III A, where the the temperature predictions were significantly inaccurate as seen in Fig. 5a. Although the general trend of temperature rise after excitation was predicted correctly by the CNN, the level of inaccuracy deemed this approach inappropriate for providing a reliable automated model replacing analytical calculations. For the model to be useful, its temperature predictions need to reliably be within a few percentage points of the analytical temperatures. This level of accuracy was achieved by our GAN-CNN network. This performance is notable, given the small size of the dataset. Despite the variation between the images being very little, the GAN can accurately recreate the small details of importance in each image. Fig. 10 shows the predicted temperatures of the network for the different fluences.

To further assess the generalizability of the model in this problem, we tested two scenarios, in which the test set included images with different fluences and different temperatures than the ones in the training set. For the first scenario, we tested the model on the 39 images with the lowest fluences (0.78 -2.6 mJ/cm2) while training it on the other images with the higher fluences. In the second scenario, we ordered all images according to their temperatures and used 40 images within the $u_{iso}$ range of 0.024 – 0.030 as the test set and the other images as training set. For both cases, we achieve prediction errors below 10%. The results are shown in Figs. 11 and 12, respectively.

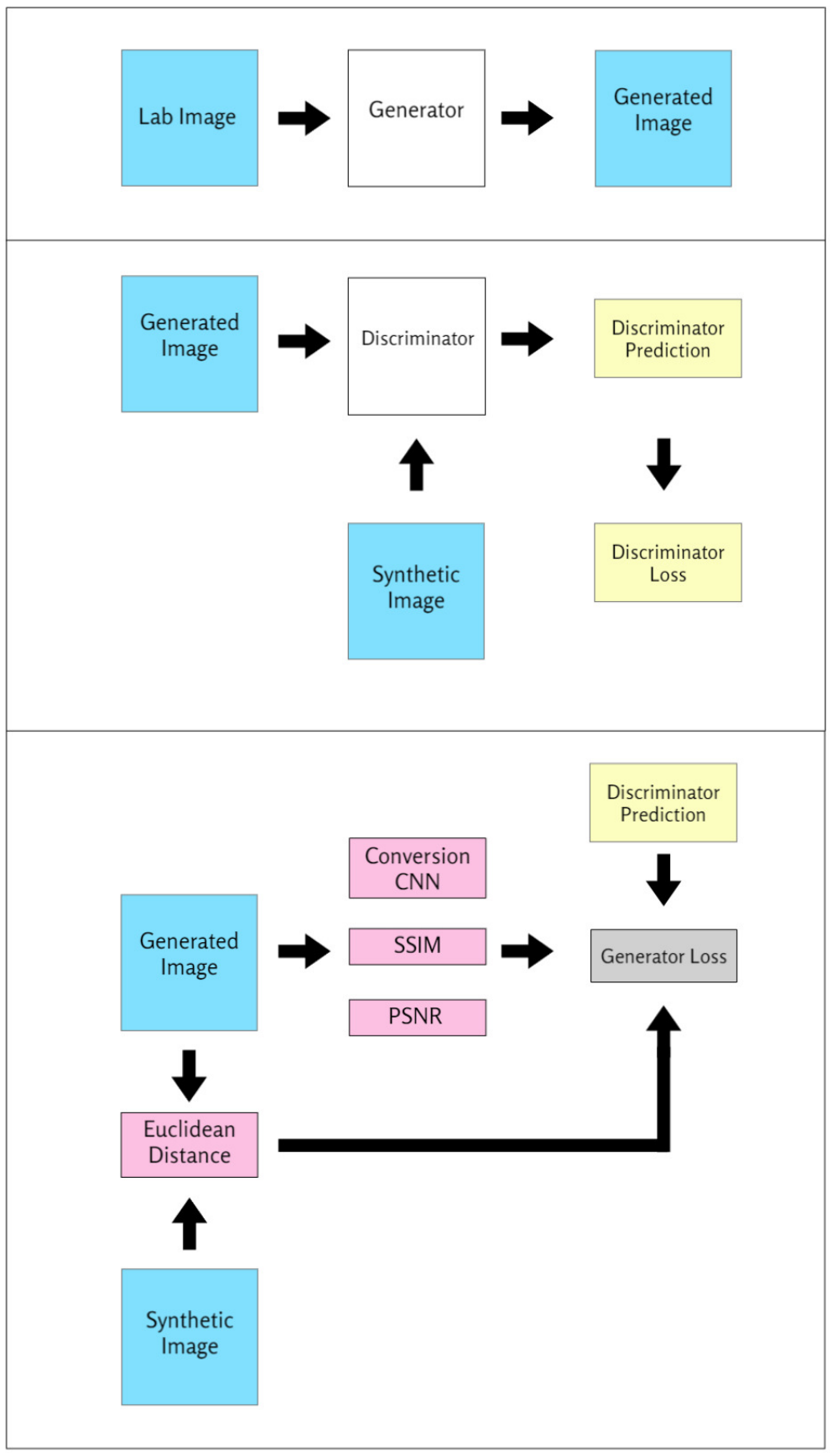

FIG. 8. Neural network schematic. (Top) The generator takes a lab image as input and generates an image emulating an ideal diffraction pattern. (Middle) The discriminator takes a generated image and a synthetic image containing an ideal diffraction pattern, with minimal processing, created via CrystalDiffract and outputs a value between 0 and 1 from which the discriminator loss is calculated. (Bottom) The total loss function of the network architecture is the sum of the squared Euclidean distance between the generated image and the synthetic image, the difference between the analytical value of $u_{iso}$ of the lab image and the value of $u_{iso}$ predicted by the CCNN, SSIM and PSNR of the generated image and the conventional generator loss which includes the discriminator predictions.

Thus, the GAN-CNN network is generalizable as it provides an image-to-image mapping between the experimental data and ideal diffraction patterns. This approach can be applied to data taken by many other instruments and in different experiments, and hence, is not limited to UED. Our data was taken over a span of several days so the network is able to discern noise profile differences from day to day and produce reliable results. However, for different instruments and experiments we believe transfer learning would be ideal to train the network quickly on new setups. We have made our code along with the weights of the parameters available on GitHub (https://github.com/dhruv-sirohi/Miller-Lab-UED) for this purpose.

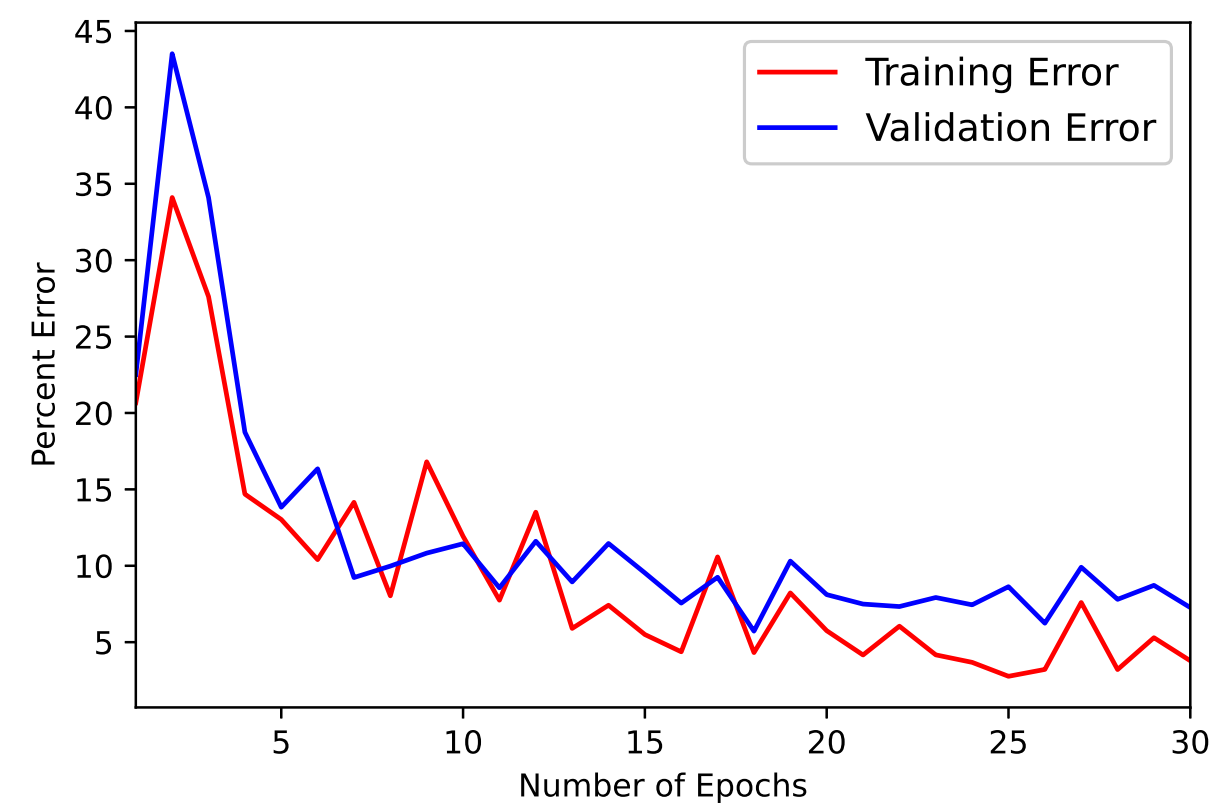

FIG. 9. Training and validation errors for temperature predictions.

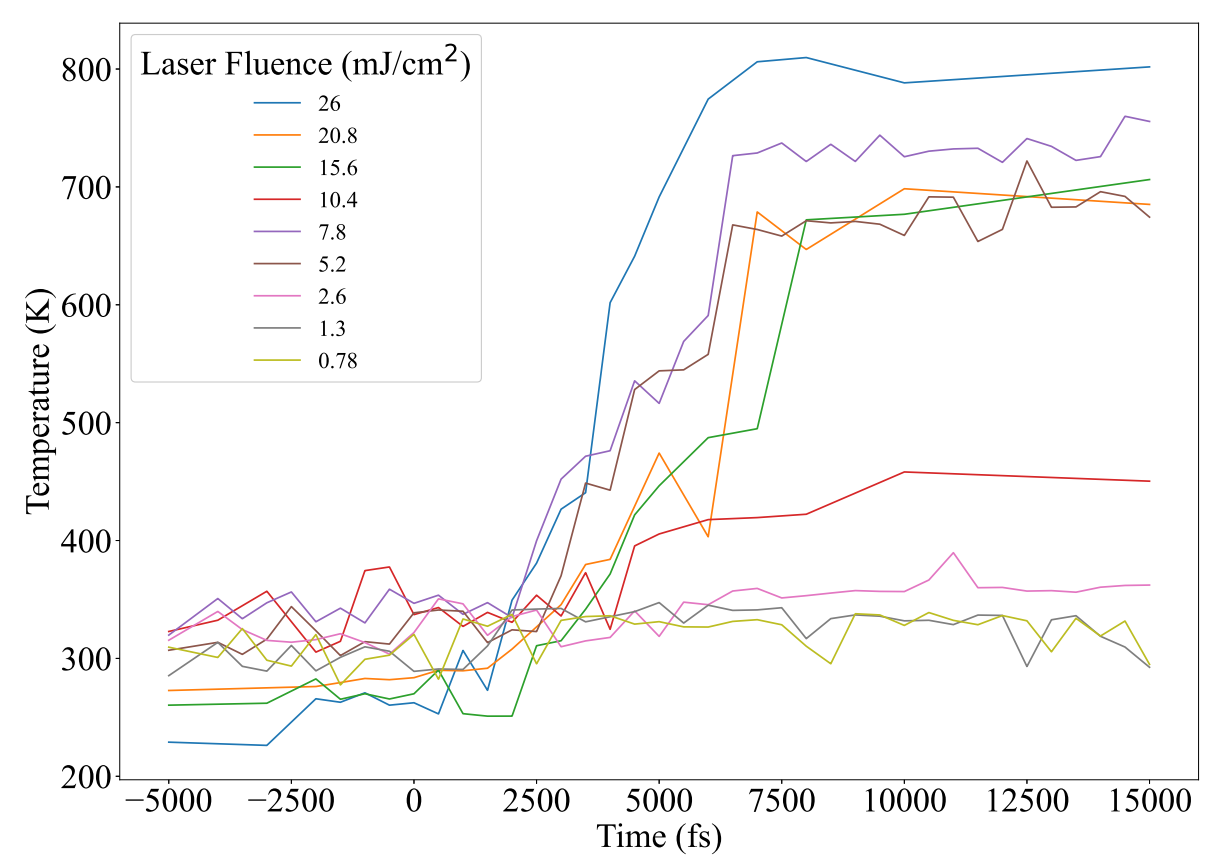

FIG. 10. The GAN-CNN neural network's predictions of the temperatures of the experimental bismuth diffraction images. The laser pulse excites the sample at t = 0 fs, where a rapid rise in temperature is observed. The temperature predictions are significantly improved as compared to the predictions of the CNN-only network in Fig. 5a.

## IV. CONCLUSION

We proposed a novel neural network that could provide considerable insights into the structural dynamics that samples undergo in typical ultrafast experiments. Our network consists of a generative adversarial network (GAN), that converts experimental diffraction images into idealized diffraction patterns, and a Conversion CNN (CCNN) that analyses the idealized images. The CCNN is solely trained on synthetic data of which an unlimited amount can be produced. We validate our approach on ultrafast electron diffraction (UED) data of polycrystalline bismuth samples that undergo thermalization upon excitation via 800 nm laser pulses. Our network predicted temperature changes with a deviation of less than 6% from the theoretically estimated values after being trained on a dataset of only 408 images.

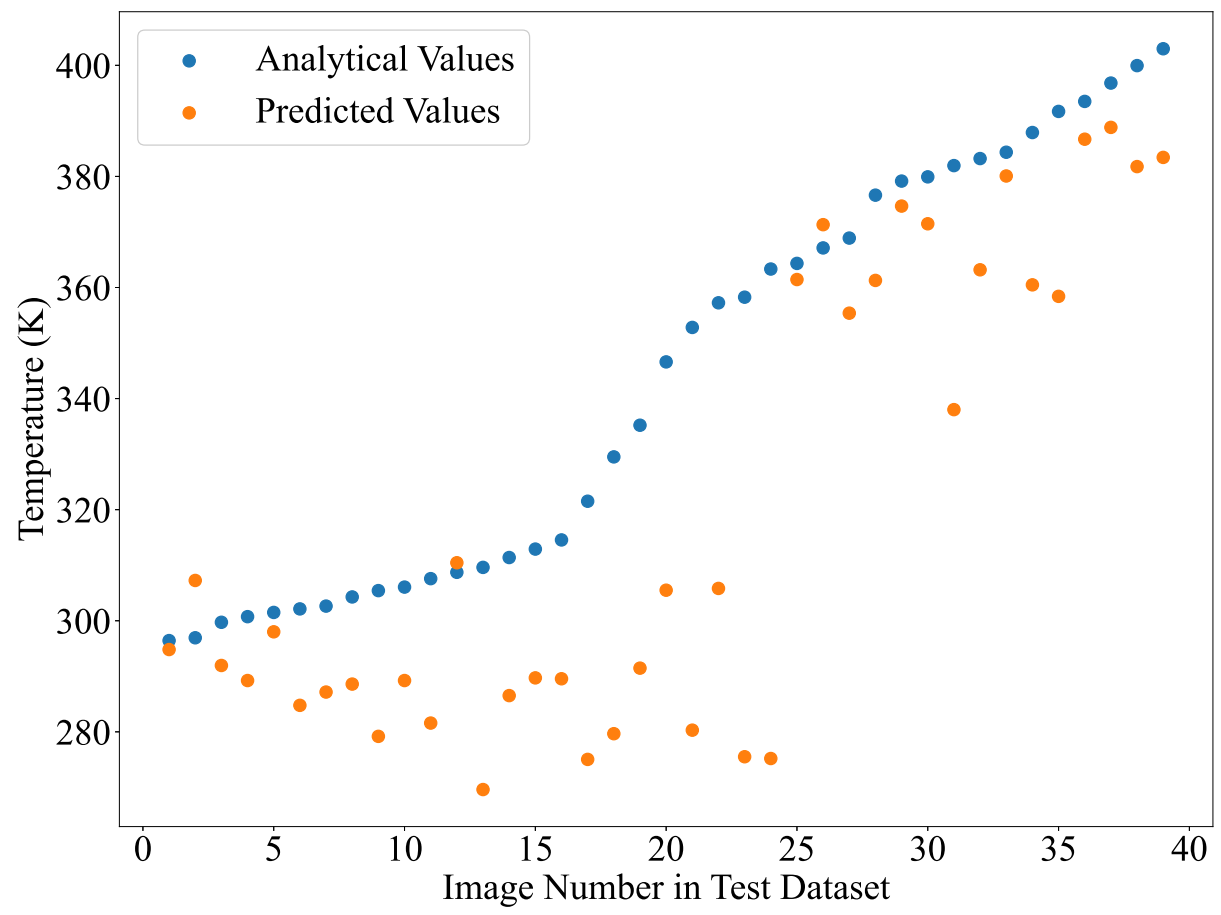

FIG. 11. Analytical and predicted temperatures for a test set containing the lowest fluences (0.78 - 2.6 mJ/cm2). The neural network was exclusively trained on images with higher fluences.

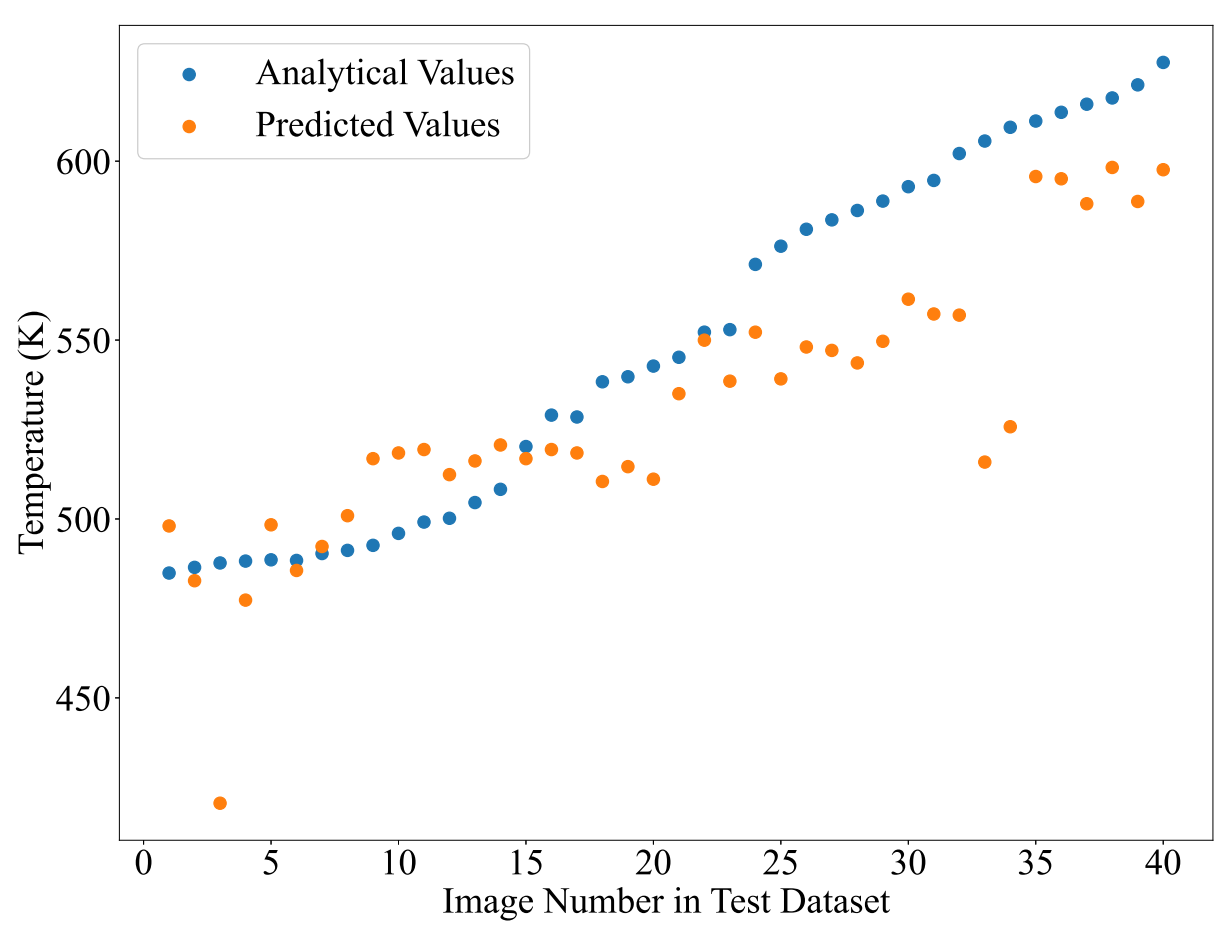

FIG. 12. Analytical and predicted temperatures for a test set with images in the $u_{iso}$ range of 0.024 – 0.030. The neural network was exclusively trained on images with $u_{iso}$-values outside this range.

As a future research direction we propose training this network in a setting with high volumes of visual data such as beam lines. Whereas we used pre- and post-excitation data for training, due to the limited dataset, we propose, as a next step, using only pre-excitation diffraction images to train the GAN. Pairing these images with the corresponding idealized diffraction patterns is straightforward, as the initial experimental conditions are known and can easily be simulated with standard software such as CrystalMaker. If data from a large number of diverse materials and initial experimental conditions is used, the GAN could be trained to "de-noise" images in new experimental conditions upon excitation. Subsequently, a Conversion CNN, trained on large data sets of synthetically created images, would be able to provide considerable insights into the dynamics the samples undergo. We hope our method would be useful for the following tasks: 1) Narrowing down the key modes to one or two modes out of a set of higher number of suspected modes. It would simply show no change in the 'non-key' modes. 2) Determining the degree by which each mode changed, e.g., if the key modes are a specific bond length and a specific rotational angle the neural network should be able to find out by how many Angstroms the bond length changed and by how many degrees the rotational angle changed at the different times.

## V. ACKNOWLEDGEMENTS

This work was supported by the National Science and Engineering Research Council of Canada (RJDM).